\documentclass[journal]{IEEEtran}
\usepackage[pdftex]{graphicx}
\usepackage{subfigure}
\usepackage{balance}
\usepackage{array}
\usepackage{times}
\usepackage{rotating}
\usepackage{multirow}
\usepackage[cmex10]{amsmath}
\usepackage{subfig}
\usepackage{tabularx}
\usepackage{float}
\restylefloat{figure or table}
\usepackage{fancyhdr}
\begin{document}

\title{Study on Scheduling Techniques for Ultra Dense Small Cell Networks}

\author{Amir H.~Jafari{$^{1}$}{$^{2}$}, David~L\'opez-P\'erez{$^{2}$}, Ming Ding{$^{3}$}, Jie~Zhang{$^{1}$}\\
{$~^{1}$}Dept. of Electronic \& Electrical Engineering, University of Sheffield, Sheffield S1 3JD, UK\\
{$~^{2}$}Bell Laboratories Alcatel-Lucent, Dublin, Ireland\\
{$~^{3}$}NICTA, Sydney, Australia\\

\texttt{a.jafari@sheffield.ac.uk}}
\maketitle

\thispagestyle{fancy}

\maketitle

\begin{abstract}

The most promising approach to enhance network capacity for the next generation of wireless cellular networks (5G) is densification, 
which benefits from the extensive spatial reuse of the spectrum and the reduced distance between transmitters and receivers. 
In this paper, we examine the performance of different schedulers in ultra dense small cell deployments. 
Due to the stronger line of sight (LOS) at low inter-site distances (ISDs), 
we discuss that the Rician fading channel model is more suitable to study network performance than the Rayleigh one, 
and model the Rician K factor as a function of distance between the user equipment (UE) and its serving base station (BS). 
We also construct a cross-correlation shadowing model that takes into account the ISD, 
and finally investigate potential multi-user diversity gains in ultra dense small cell deployments by comparing the performances of proportional fair (PF) and round robin (RR) schedulers. 
Our study shows that as network becomes denser, 
the LOS component starts to dominate the path loss model which significantly increases the interference. 
Simulation results also show that multi-user diversity is considerably reduced at low ISDs, 
and thus the PF scheduling gain over the RR one is small, around 10\% in terms of cell throughput. 
As a result, the RR scheduling may be preferred for dense small cell deployments due to its simplicity. 
Despite both the interference aggravation as well as the multi-user diversity loss, 
network densification is still worth it from a capacity view point.

\end{abstract}


\section{INTRODUCTION}

Technologies that will bring the fifth generation of cellular networks (5G) to reality are currently under investigation,
with the following three,
being the main approaches to manage the data deluge:
\begin{itemize}
\item
use of higher frequency carriers to benefit from more bandwidth,
\item
use of more antennas to obtain higher spectral efficiencies, and
\item
network densification through heterogeneous networks (HetNets) to exploit spatial reuse.
\end{itemize}

Among these three approaches,
network densification is envisioned to be the key solution to meet users' demands
since it has the potential to linearly increase the network capacity with the number of deployed cells.

In a HetNet, low power small cells co-exist with high power macrocells.
Low power small cells aim to satisfy the traffic demands at hotspot locations,
while high power macrocells provide an umbrella coverage to support user equipments (UEs) with high mobility.

In order to provide an efficient mobility management and avoid UE's constant handovers among base stations (BSs), 
network densification is not appropriate for macrocell BSs.
It is more feasible to consider network densification only for small cell BSs, 
which benefits from a lower cost and deployment flexibility due to their reduced form factor~\cite{6815892}~\cite{6525591}.
Small cells can be placed in strategic locations to leverage the current infrastructure,
while taking into account UE distributions, traffic demand and radio propagation conditions.
Moreover, dense small cell networks can operate in a different frequency band than macrocells ones,
significantly mitigating interference.

Small cell  network densification brings about significant benefits:
\begin{itemize}
\item
Results in an intense spatial reuse,
\item
Allocates larger shares of the available spectrum to UEs
due to reduced number of UEs served per small cell BS, 
and
\item
Brings down the path loss by decreasing the distance between the small cell BSs and the UEs.
\end{itemize}
However, despite of their benefits,
small cell network densification also opens up new research questions,
for example, 
in terms of radio resource management. 

Scheduling has been conceived as an effective technique to efficiently use the available spectrum and improve network throughput 
in macrocell scenarios with a large number of UEs per macrocell BS.
In more detail, proportional fair (PF) scheduler is used as an appealing scheduling technique 
that offers a good trade-off between maximising throughput and improving fairness among UEs with diverse channel conditions.
However, the gains of PF may be limited in dense small cell networks, 
partly because the number of UEs per small cell BS is considerably reduced in comparison to macrocell ones,
and partly because UEs may not experience very different channel conditions on different subcarriers due to the dominance of LOS propagation as UEs may be really close to their serving BS.
These changes give rise to the question of whether the PF scheduling is as efficient for dense small cell networks 
as it is for macrocell scenarios,
or if it can be substituted by schedulers of lower complexity.

This paper analyses different scheduler types under different  densification levels,
and analyses some  fundamental tradeoffs of network densification.
The smaller the cell size, 
the closer the UE is to its serving BS -- reducing the path loss,
but the stronger the LOS -- reducing the multi-user diversity.
Furthermore, decreasing the cell size will not only diminish the multi-user diversity,
but also will increase the interference due to the dominance of LOS.
As a result, it is necessary to take a systematic view towards the tradeoffs of network densification.

The rest of this paper is organised as follows.
In section~\ref{sec:fading},
we discuss the Rician fading model and model the Rician~K factor considering the LOS probability and the distance between the UE and its serving BS.
Moreover, as we consider different cell sizes,
we also propose an ISD dependent shadow fading cross correlation coefficient.
In section~\ref{sec:schedul},
different scheduling algorithms are presented.
In Section~\ref{sec:sim},
we analyse the simulation results for different scheduling algorithms in a range of dense small cell deployments,
and examine system performance in response to different ISDs.
In Section~\ref{sec:con}, the conclusions are drawn.

\section{Channel Fading Models in a Dense Small Cell Network}
\label{sec:fading}

\subsection{Small Scale Multi-Path Fast Fading Model}

A radio signal may travel along different paths and therefore multiple copies of the signal may arrive at the receiver at distinct time instants and with different phases causing multi-path combination,
which could be constructive or destructive.
In this section, we model the multi-path fast fading as a function of distance between the UE and the BS, incorporating the probability of LOS.

Considering the cell size and the relative proximity of UEs to their serving BSs,
there is a high probability of LOS in dense small cell networks,
which indicates that Rician fading channel models may be more appropriate than Rayleigh ones to model multi-path channels in this type of deployment.
The Rician fading model considers a dominant, non-fluctuating strong path in addition to a number of reflections and scatterings,
referred to as LOS and non-LOS (NLOS) components, respectively~\cite{HetNetbook}.
Using more realistic fading models based on measurements,
in which the delay spread is considered,
will help to understand the impact of different types of environments on the performance results too,
and this is left for our future work.

In this paper, the probability distribution function (PDF) of Rician fading is given as
\begin{equation}
\vspace{-0.5mm}
\begin{split}
f(x) =& \left[ \frac{2 (K+1)x}{\gamma} \exp(-K-\frac{(K+1){x^2}}{\gamma}) \right. \\
  & \left. I_{0}(2 \sqrt{\frac{K (K+1)}{\gamma}} x) \right],
\end{split}
\label{eq:rician}
\end{equation}
where $\gamma$ refers to the total power in LOS and NLOS components,
and $I_{0}$ is the first kind $0^{th}$ order modified Bessel function.
The Rician K factor denoted by $K$ is the ratio of power distribution in the specular LOS to the NLOS multi-path components,
and ranges between $0$ and $\infty$, 
with both extremes corresponding to the Rayleigh channel and the non-fading channel, respectively. 
The Rayleigh fading occurs when there is no dominant LOS path.

\begin{figure*}
\centering
\subfigure[Probability of LOS versus UE-Cell distance.]{\includegraphics[scale=0.405]{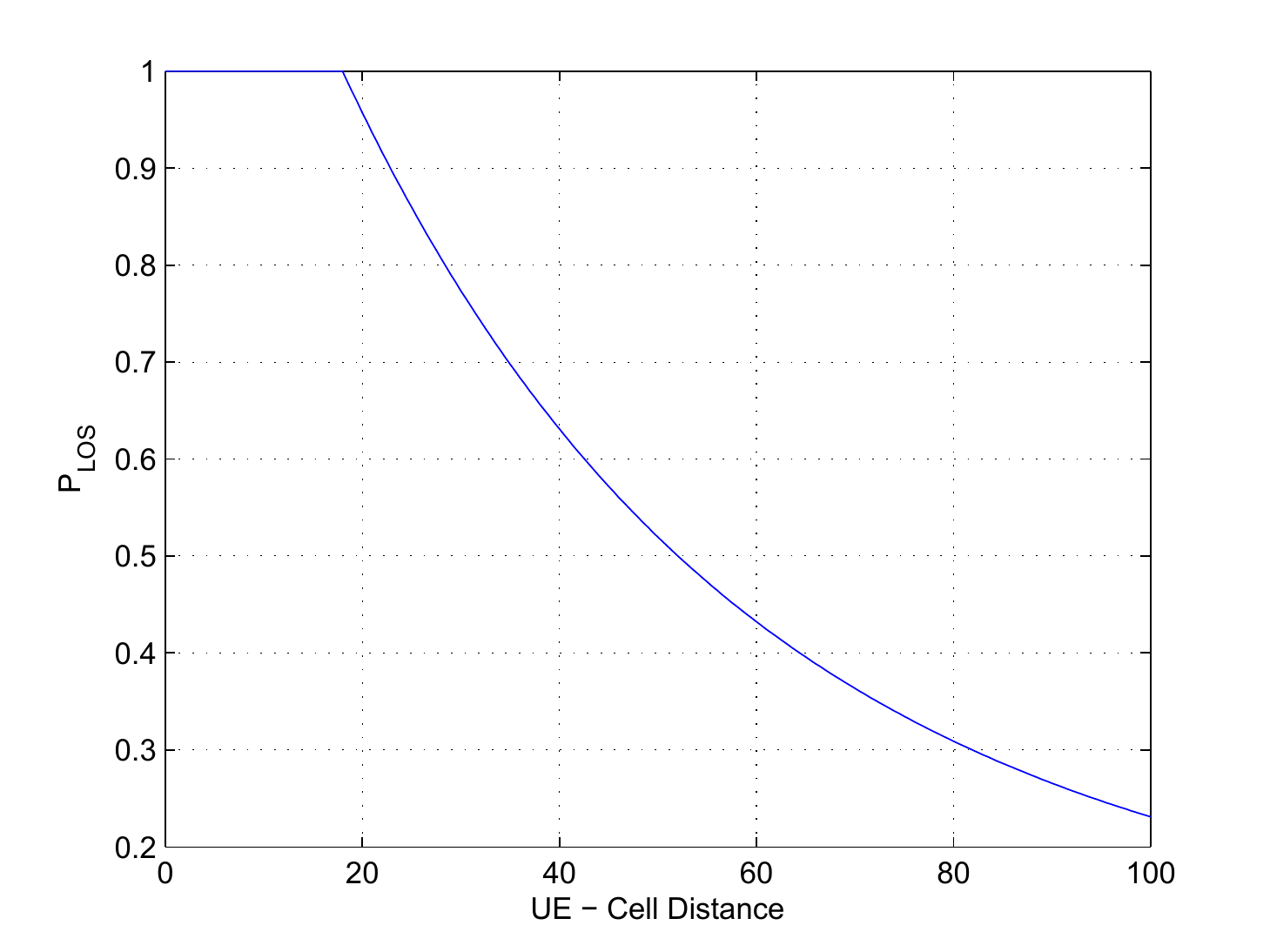}}\quad
\subfigure[Rician K factor versus UE-Cell distance.]{\includegraphics[scale=0.405]{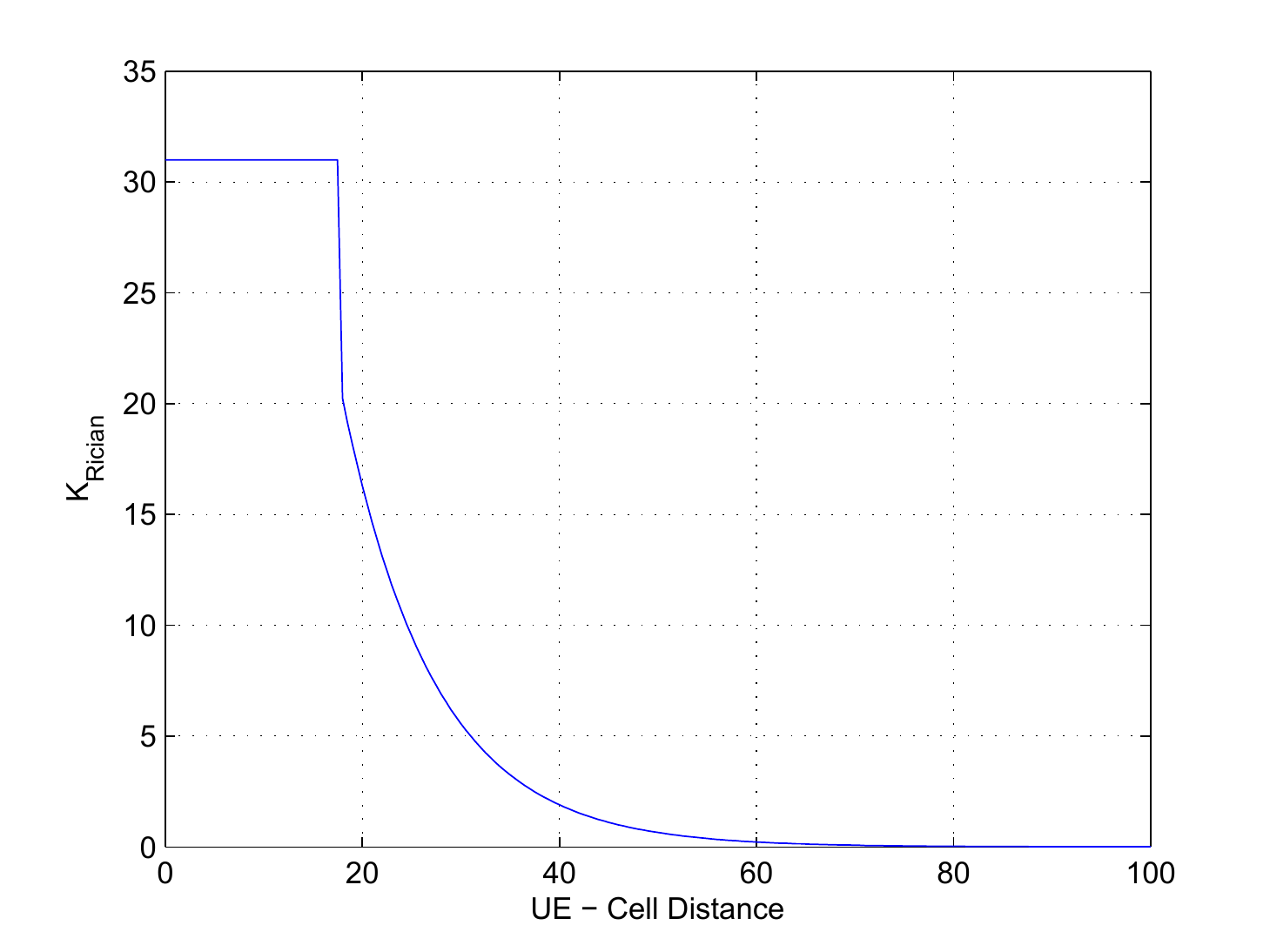}}\quad
\caption{One-to-One correspondence between $P_{\rm LOS}$ and the $K$ factor.}
\label{fig:los_rician}
\vspace{-0.6cm}
\end{figure*}

In this study, a new model where the K factor is derived according to the probability of LOS is proposed. 
The probability of LOS as a function of distance for micro urban scenarios is given as
\begin{equation}
\vspace{-0.5mm}
P_{\rm LOS}= \min(\frac{18}{d},1) \times (1-e^{\frac{-d}{36}}) + e^{\frac{-d}{36}},
\label{eq:plos}
\end{equation}
where $d$ is the distance between the UE and its serving BS~\cite{3gpp}.
According to this model,
within 18\,m from the BS,
the $P_{\rm LOS}$ is equal to $1$.
UEs that are positioned up to 18\,m from the BS location have a guaranteed strong LOS component.

In order to comply with the $P_{\rm LOS}$ of 1,
the value of 32 ($\sim$15 dB) is assigned to the K factor to secure the existence of a strong LOS component within the LOS zone.
This value is specified since it results in the standard deviation of the Rician fading being smaller than 0.5~dB, flat fading.
As UEs locate further away from the BS,
the $P_{\rm LOS}$ exponentially decays and so should the K factor due to the proposed one-to-one correspondence. Accordingly, the K factor is approximated as $\rm K = \frac{\rm P_{\rm LOS}}{\rm P_{\rm NLOS}}$ where $P_{\rm NLOS}$ is equal to $1 - P_{\rm LOS}$ and hence for the Out-of-LOS zone, the K factor is modelled by the exponentially decaying function shown in (\ref{eq:los}). 
\begin{equation}
\vspace{-0.5mm}
  \rm K=\begin{cases}
      32 & \text{if $d<18 m$} \\
     140.10 \times \exp(-0.107 \times d) & \text{otherwise},
  \end{cases}
  \label{eq:los}
\end{equation}

This is interpreted as a distance dependant transition from Rician to Rayleigh fading 
for UEs that are located further away from their BSs
where the LOS component gradually fades.
Fig.~\ref{fig:los_rician} shows the derived K factor and the corresponding $P_{\rm LOS}$.
It is important to note that this new model can be further extended and calibrated with measurements.

\subsection{Large Scale Shadow Fading Model}

Due to multiple UEs and BSs,
the shadow fading model has to consider the spatial auto and cross correlation properties~\cite{1651489}.
Cross-correlation shadow fading refers to the situation where a given UE may see similar shadow fadings from different BSs deployed near each other.
It is typically assumed,
mostly within the 3GPP studies,
that there is a 50\% cross-correlation among different sites' shadow fadings~\cite{1651489}~\cite{3gpp}.
However, this is idealistic and the degree of cross-correlation should be ruled by the distance among BSs.
In (\ref{eq:shad}), 
a new cross-correlation shadowing model which is a function of ISD is proposed,
\begin{equation}
\vspace{-0.5mm}
\rho_{\rm cross}(\Delta x) = \min \left[\sqrt{0.5^2 + \exp^{2}(-\frac{\rm ISD}{\rm d_{cor}})},1\right],
\label{eq:shad}
\end{equation}
where $d_{cor}$ is the decorrelation distance and is referred to the distance where the cross-correlation coefficient drops to 0.5.
In this model, lower values of ISD will result in higher spatial correlation between shadow fadings of nearby BSs.

\section{Scheduling Algorithms for Small Cells}
\label{sec:schedul}

In long term evolution (LTE), a resource block (RB) refers to the basic time/frequency  scheduling resource unit to which a UE can be allocated.
Each RB expands 180 KHz in the frequency domain and has a duration of  1\,ms in the time domain.
The RB consists of 12 subcarriers of 15\,KHz and its 1\,ms transmission time interval (TTI) is referred as sub-frame.

Unlike other diversity techniques that aim to average the signal variations to mitigate the destructive impact of multi-path fast fading,
multi-user diversity, 
also known as channel sensitive scheduling, 
takes advantage of multi-path fading
by allocating to each RB the UE that has the best channel conditions to enhance network performance~\cite{LTE}.
Such type of scheduling leads to multi-user diversity gains,
which have been shown to roughly follow a double logarithm scaling law in terms of capacity with regard to the number of UEs per BS 
for macro cell scenarios~\cite{4063519}.
It is important to note that in order to aid the channel sensitive scheduling and exploit multi-user diversity gains,
UEs need to report downlink channel quality indicators (CQI) back to their serving BSs,
which allow the scheduler to asses the UE channel quality and perform the scheduling according to an specified metric.

In a network with $N$ UEs,
each UE may undergo varying channel conditions
where better channel quality generally refers to higher signal quality and higher throughput.
Sharing the resources fairly between UEs experiencing different channel qualities is a challenging task~\cite{4146798}.

\emph{Opportunistic schedulers} selects the UE with the best channel quality at each time/frequency resource,
aiming to solely maximise the overall throughput,
whereas \emph{Round Robin (RR)} schedulers treat the UEs equally regardless of their channel qualities,
giving the same amount of time/frequency resources to all UEs. 
The former scheduler can increase system throughput remarkably compared to the latter at the expense of fairness, 
since UEs with relatively bad channel qualities may be never scheduled~\cite{4489366}.
Different types of RR schedulers are summarised in Table~\ref{tab:com}.

\emph{PF schedulers} exploit multi-user diversity based on UEs CQI,
attempting to maximise the throughput while simultaneously forcing a degree of fairness in serving all the UEs.
The PF scheduling metric basically aims at weighting the UE's potential instantaneous performance by its average performance,
and this process consists of three stages.
In the first stage, according to buffer information,
the schedulable set of UEs is specified.
The second stage is the time domain scheduling,
which is in charge of reinforcing fairness and  selecting the $N_{max}$ UEs that will be input to the frequency domain scheduler.
The last stage corresponds to the frequency domain scheduling,
i.e., allocation of UEs to RBs.
The complexity of the frequency domain scheduler highly depends on the number of its input UEs~\cite{4526113}~\cite{851593},
and thus the time domain scheduler has a major impact on the complexity of the frequency domain scheduler.

A PF scheduling metric in time domain can be defined as
\begin{equation}
\vspace{-0.5mm}
M_{\rm PF-TD} = \frac{\hat{D}[n]}{R[n]},
\label{eq:td}
\end{equation}
where $R[n]$ and $\hat{D}[n]$ are the past average throughput and potential instantaneous throughput of $n^{th}$ UE, respectively~\cite{4526113}.
The past average throughput can be computed using a moving average as
\begin{equation}
\vspace{-0.5mm}
R_{i}(t+1) = (1-\frac{1}{T_{c}}) R_{i}(t) + \frac{1}{T_{c}} \times r_{i}(t),
\label{eq:exp}
\vspace{-1.4mm}
\end{equation}
where $\rm T_{c}$ is the length of the moving average window and should be larger than the time elapsed between multiple schedules of the individual UE,
and $r_{i}$ is the current data rate of the serving UE.
It is worth noting that the current data rate of a non-serving UE is considered to be zero.
The UEs will be ranked constantly according to the metric in (\ref{eq:td}),
and the $N_{max}$ UEs with maximum preference are passed on to the frequency domain scheduler.

\begin{figure}[t]
\centering
\includegraphics[scale=0.37]{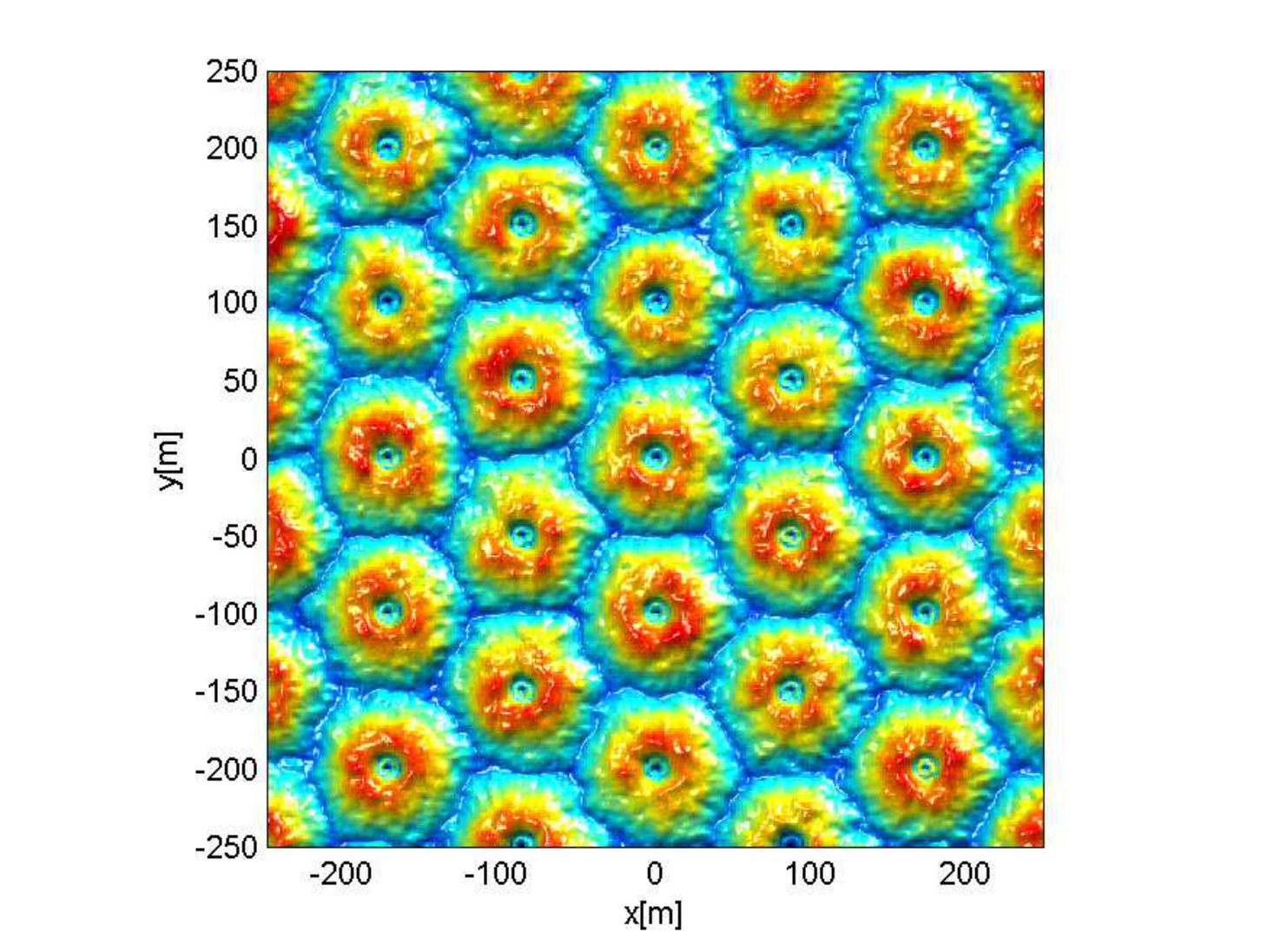}
\caption{Multi-tier hexagonal grid of small cells.}
\label{fig:tier}
\vspace{-0.7cm}
\end{figure}
%[David]: This can be taken out if needed.

A PF scheduling metric in frequency domain can be defined as
\begin{equation}
\vspace{-0.5mm}
M_{\rm PF-FD} = \frac{SINR[n,k,t]}{\sum\limits_{k = 1}^{\rm N_{RB}} SINR[n,k,t]},
\label{eq:pf}
\end{equation}
where the numerator is the signal-to-interference-plus-noise ratio (SINR) of the $n^{th}$ UE on the $k^{th}$ RB
and the denominator is the sum of the $n^{th}$ UE SINRs over all RBs,
which represents its average channel quality at sub-frame $t$~\cite{4526113}.
The UEs will be ranked constantly according to the metric in (\ref{eq:pf}),
and the one with maximum preference is selected to be served on $k^{th}$ RB at each sub-frame. 

The SINR of $n^{th}$ UE over $k^{th}$ RB at sub-frame $t$ is modelled as 
\begin{equation}
\vspace{-0.5mm}
SINR_{(n,k,t)} = \frac{P_{(n,k,t,i)} h_{(n,k,t,i)}}{{\sum\limits_{j = 1 \hspace{0.1cm}\& \hspace{0.1cm} j\neq i}^{\rm N_{BS}} P_{(n,k,t,j)} h_{(n,k,t,j)}} + N_{0}},
\label{eq:snr}
\end{equation}
where $P$ is the BS transmit power over the $k^{th}$ RB,
$i$, $j$ and $N_{BS}$ are the index of the serving BS, index of interfering BSs, and the total number of BSs, respectively, 
$h_{(n,k,t,m)}$ is the total channel gain between the $n^{th}$ UE and the $m^{th}$ cell, 
comprising antenna gain, path loss gain, shadow fading gain and multi-path fast fading gain 
and $N_{0}$ is the noise power.

\section{SIMULATION RESULTS}
\label{sec:sim}

We consider a multi-tier hexagonal layout of small cell BSs in a $500m \times 500m$ scenario.
Different ISDs are considered to study how system performance responds to different degrees of network densification.
Macrocell BSs operate in a different frequency band.
The central small cell is designated as the serving cell where the UEs are uniformly distributed and the rest act as interferers.
Antenna gain, path loss, lognormal shadowing and multi-path Rician fast fading are included in SINR computation.
RR and PF schedulers are then applied to observe how RB allocation mechanisms affect the UE and cell throughput.
The number of UEs served by the serving cell is also varied from a single UE scenario to multiple ones,
with the latter aiming to study the effectiveness of scheduling in exploiting multi-user diversity under different UE loads.
Since Rayleigh fading models are commonly used in the literature,
comparisons between the more realistic Rician model (due to LOS presence) and the Rayleigh one are also made in terms of network performance.
Table~\ref{tab:par} summarises the simulation setting.

\begin{table}[t]
\centering
\renewcommand{\arraystretch}{1.4}
\caption{System Simulation Parameters}
\label{tab:par}
\scalebox{0.85}{
\centering
\begin{tabular}{|>{\centering\arraybackslash}p{4.1cm}|>{\centering\arraybackslash}p{1.8cm}|}
\hline \textbf{Parameter} &  \textbf{Setting} \\ \hline
\hline
Carrier Frequency (GHz)             		& 2      \\ \hline
Transmission Bandwidth (MHz)             	& 10     \\ \hline
Number of RBs       				& 50      \\ \hline
Sub-Frame Duration (ms)                 		& 1     \\ \hline
Numer of Simulated Subframes     		& 100      \\ \hline
$\rm T_{c}$ (ms)                 			& 4     \\ \hline
$\rm \alpha$ (deg)                  			& 8.045     \\ \hline
$\rm h_{UE}$ (m)                 			& 1.5     \\ \hline
Shadow Fading STD (dB)             		& 4 \\ \hline
$\rm d_{\rm cor}$ (m)             			& 20 \\ \hline
$\rm N_{max}$						& $\rm N_{UE}/2$ \\ \hline
\end{tabular}}
\vspace{-0.5cm}
\end{table}

The small cell BS antenna considered is a micro dipole array currently used in small cell BS products~\cite{6362524}.
The antenna height and consequently antenna gain depends on the ISD, 
which can be calculated as
\begin{equation}
\vspace{-0.5mm}
\rm h_{\rm SCBS} = \rm \frac{ISD}{\sqrt{3}} \times \tan(\alpha) + h_{\rm UE},
\label{eq:height}
\end{equation}
where $\alpha$ is the angle between the -3~dB antenna beam ray and the horizontal axis
and $\rm h_{UE}$ is the UE antenna height.
The path loss model used is the microcell urban model defined in~\cite{3gpp},
which includes LOS and NLOS components,
as introduced earlier.

%Number of tiers
Since statistics are collected only from the central small cell BS, 
to determine the required number of interfering tiers,
we performed simulations with up to 3 additional interfering tiers around the central cell respectively corresponding to 6, 18 and 36 interfering cells in the hexagonal layout.
The simulation results show that due to increased distance between BSs and UEs and the low shadow fading standard deviation in this case,
the interference by second and third tiers is negligible and so the addition of extra tiers has very minimal impact on UE SINR.
In light of these results, only one interfering tier around the central cell is considered in the sequel. 
This is in line with~\cite{4525984}.

\begin{figure}
\centering
\includegraphics[scale=0.5]{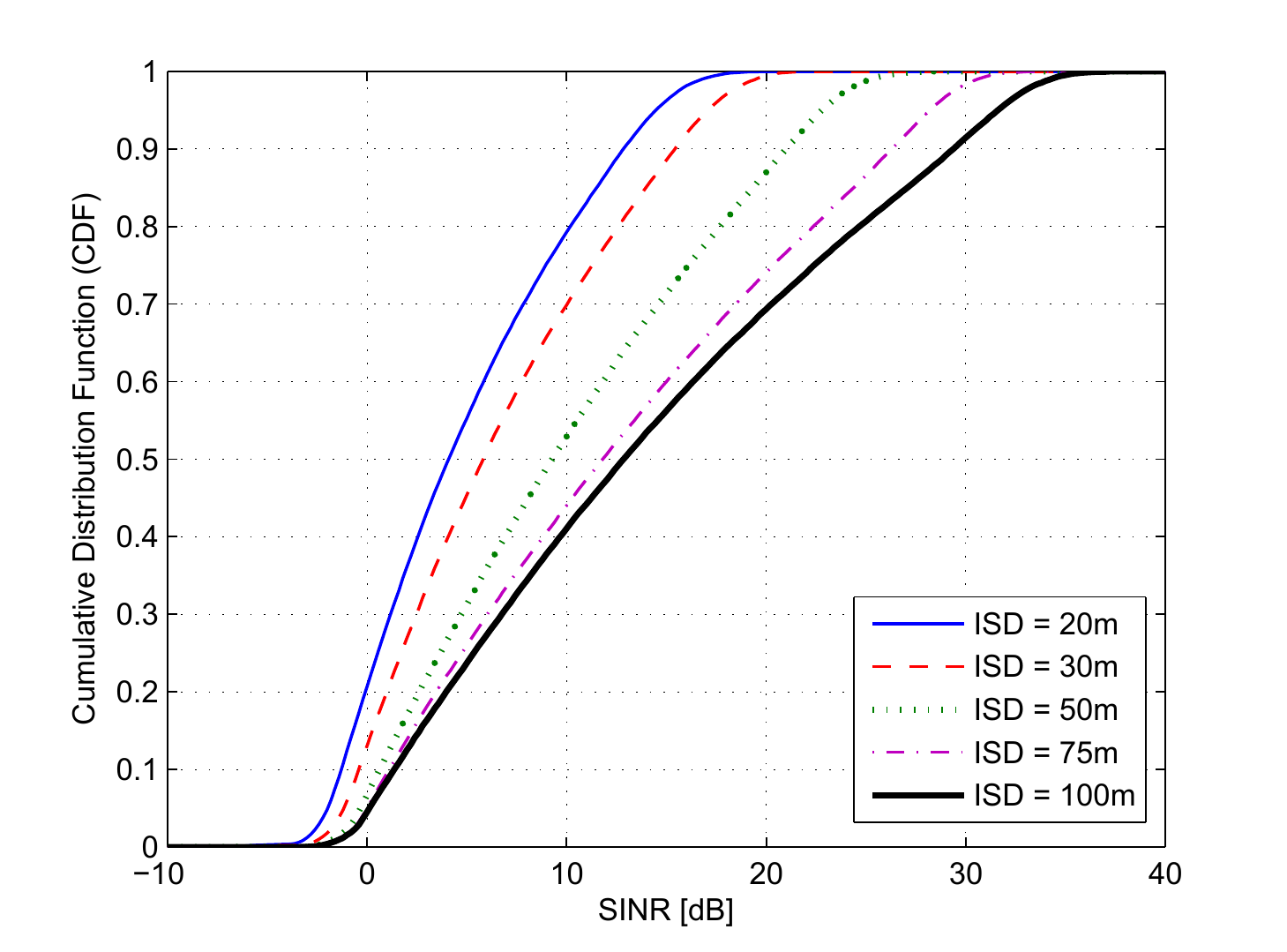}
\caption{CDF of SINR for different ISD.}
\label{fig:sinr_cdf}
\vspace{-0.6cm}
\end{figure}

%SINR distribution
Fig.~\ref{fig:sinr_cdf} shows the impact of the cell size on the cumulative distribution function (CDF) of UE SINR.
As having pointed earlier,
as the network becomes denser, the ISD is reduced and the LOS component starts to dominate the path loss model for both the carrier and the interfering signals.
However, it is important to note that in macrocell scenarios
while the carrier signal may be subject to LOS depending on the distance between the UE and its serving BS,
the interfering signal usually was not subject to LOS due to the large distance between macro BSs.
With the smaller cell sizes,
LOS starts to dominate the interfering signal too and this degrades the UE SINR with the ISD which lowers the UE and cell throughputs.

\subsection{Impact of Rician K Factor}

\begin{figure}
\centering
\includegraphics[scale=0.545]{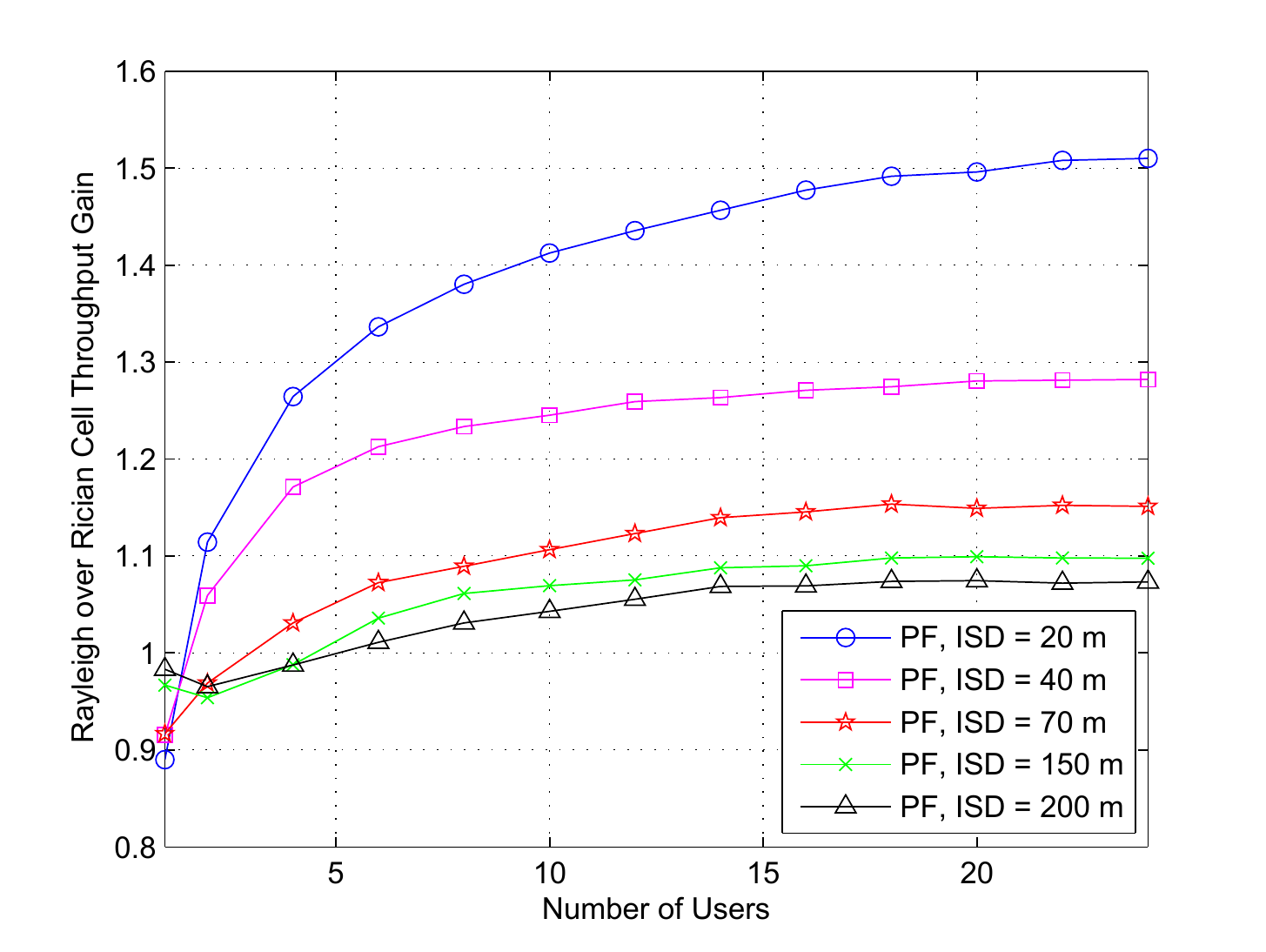}
\caption{Rayleigh over Rician Fading Gain with PF scheduling.}
\label{fig:gain}
\vspace{-0.4cm}
\end{figure}

It was discussed that Rician fading is more appropriate to model the channel. However, in order to study the impact of the Rician K factor on the scheduler performance, the Rayleigh fading channel model for which the Rician K factor is zero, is also inspected.
Fig.~\ref{fig:gain} shows the ratio of cell throughput under Rayleigh channel model to Rician one. The UE-Cell distance varies when using various ISDs and from~(\ref{eq:los}), this results in different K factors. 
Moreover, for a given ISD, 
the more UEs being served, the higher the Rayleigh gain,
since under the Rayleigh model there is more fluctuations in channel conditions and so will be more multi-user diversity.
For example, for an ISD of 20\,m, 
the Rayleigh model boosts the cell throughput by 1.3x when having 5 UEs,
while the boost is 1.41x when having 10 UEs.
However, according to (\ref{eq:los}), 
as the ISD increases the K factor decreases and  the Rician channel model becomes more a Rayleigh one,
thus diminishing the Rayleigh gain.
Fig.~\ref{fig:gain} shows that while serving 5 UEs the Rayleigh over Rician gain drops by nearly 23\% when ISD is increased from 20 to 70\,m.
Since the Rayleigh model is unrealistic for small ISD due to LOS presence and because it results in very optimistic performances due to the over estimation of multi-user diversity,
in the sequel, we adopt the Rician channel model.

%%%Reviewed until here 
\subsection{Performance Evaluation of Scheduling Algorithms}

In this section, the performance of RR and PF schedulers are compared under the Rician channel model.
Since all RR schedulers had similar performance with a $\sim$ 2\% variance,
only the one that provided the best performance (RR~4) is considered in the discussion for the sake of presentation.

\begin{figure}[t]
\centering
\includegraphics[scale=0.545]{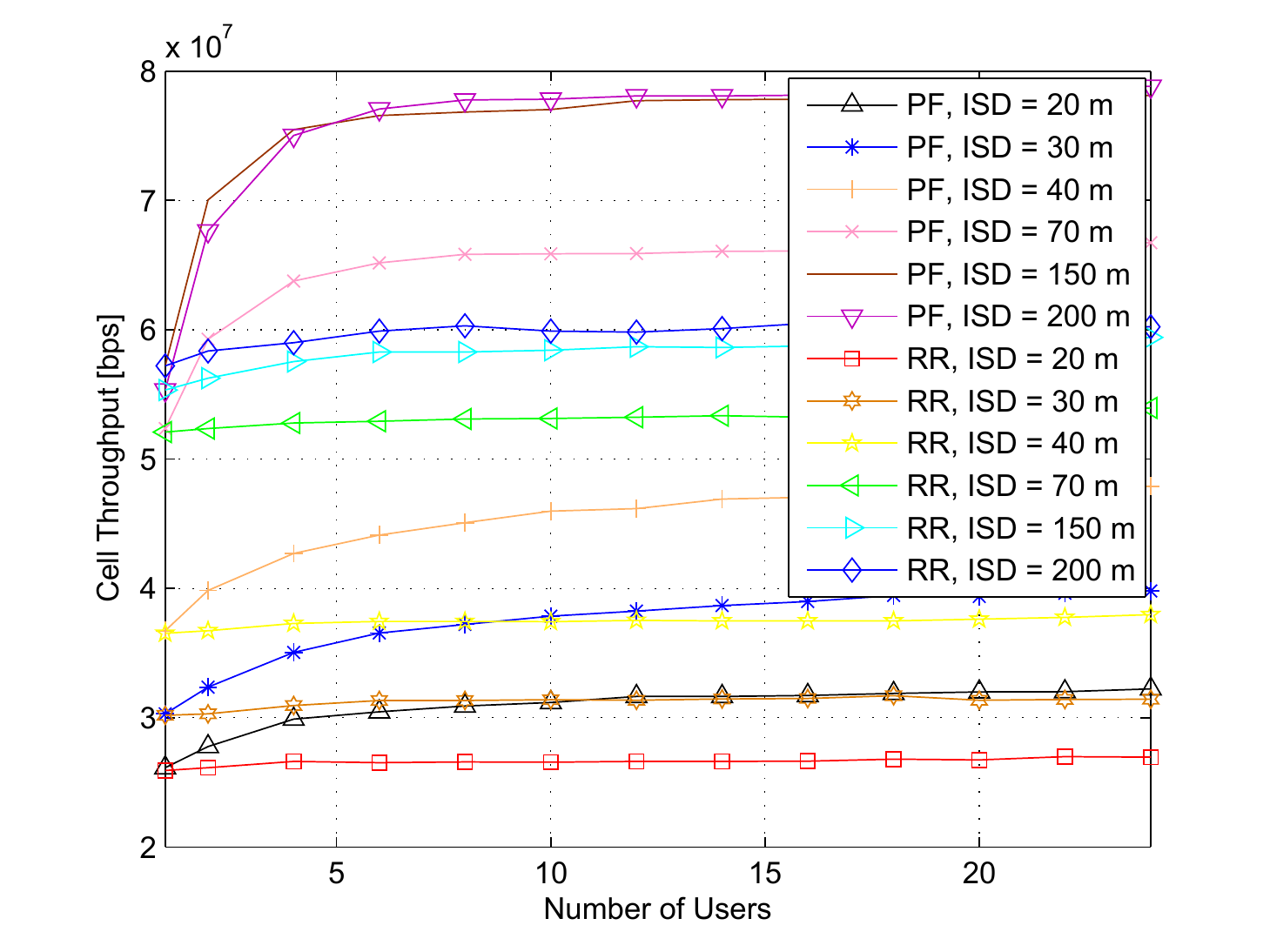}
\caption{Mean cell throughput for various ISD with different schedulers.}
\label{fig:cell_ov_thr}
\vspace{-0.6cm}
\end{figure}

Fig.~\ref{fig:cell_ov_thr} shows the performances of RR and PF schedulers in terms of cell throughput with respect to the number of served UEs for different ISDs.
%RR
As can be seen, when using RR,
the number of served UEs per BS does not impact the cell throughput,
since RR does not take into account the UE channel quality and therefore does not take advantage of multi-user diversity.
%PF
In contrast, PF is able to benefit from  multi-user diversity,
and the cell throughput increases with the number of served UEs.
%PF limit 
However, it is important to note that PF can only exploit multi-user diversity until a given extent,
as it is analysed in the following.

In terms of number of served UEs per BS,
there is a point in which a further increase in such number does not bring any significant cell throughput gain.
For example, for an ISD of 200\,m having more than 8 UEs per cell does not noticeably increase cell throughout,
while for an ISD of 20\,m,  this number is reduced to 6 UEs.
This shows how multi-user diversity gains vanishes with network densification,
regardless of the number of UEs per BS,
due to stronger LOS propagation and less fluctuating channel conditions.

Simulation results also show how the PF scheduler starts losing its advantage in terms of cell throughput with the reduced cell size.
For a given number of UEs per BS, let's say 4, 
the PF gain over RR is about 10.5\,\%, 12.4\,\% and 21.2\,\%  for dense deployments with ISDs of 20, 40 and 150\,m, respectively.
This efficiency loss of PF with the cell size makes us wonder if it is the suitable scheduler in dense small cell deployments.

\begin{figure}[t]
\centering
\includegraphics[scale=0.545]{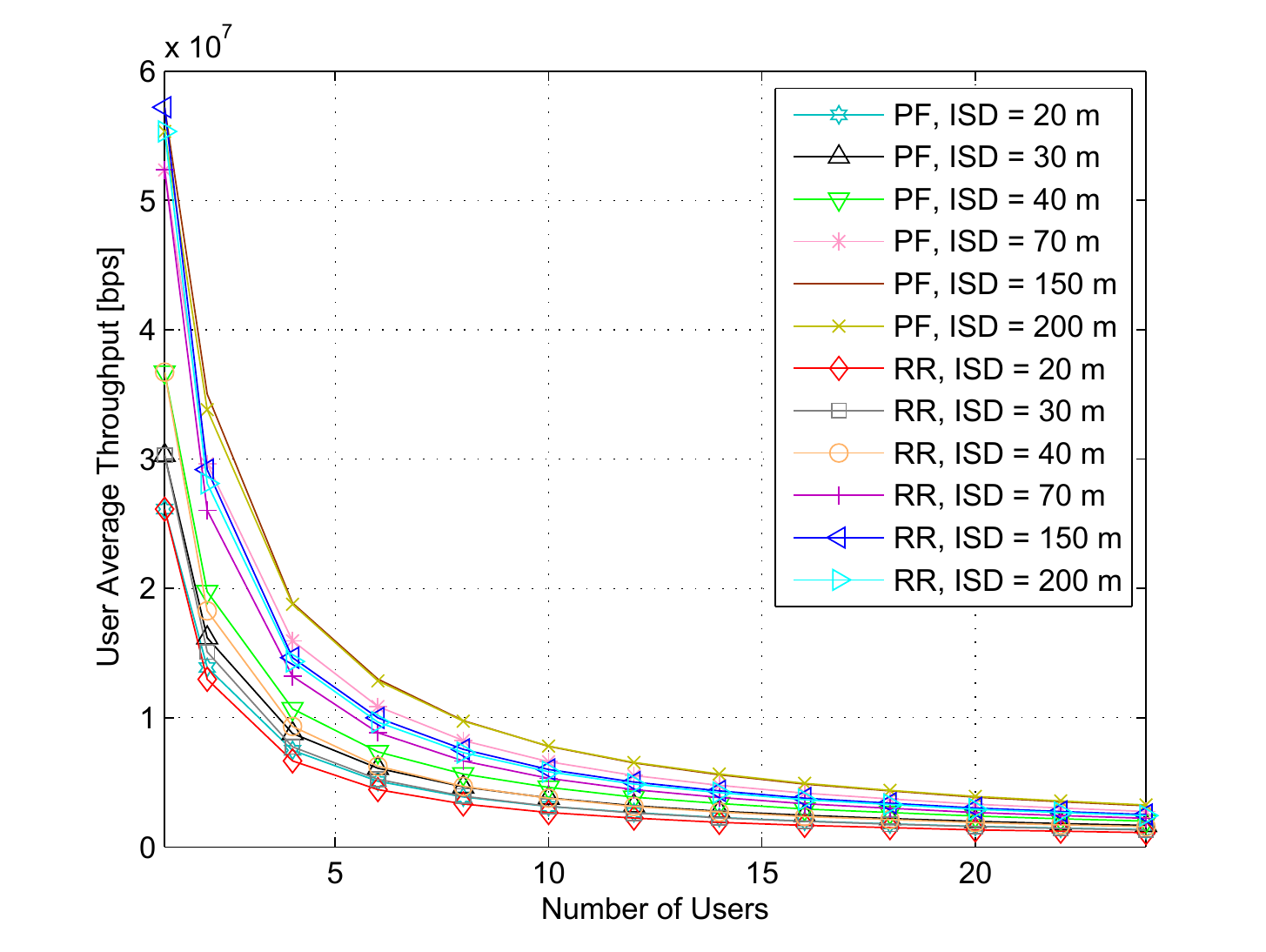}
\caption{UE mean throughput for different ISDs.}
\label{fig:ue_thr}
\vspace{-0.28cm}
\end{figure}

Shrinking the cell size not only reduces cell throughput but also reduces UE throughput.
Applying the PF scheduler,
Fig.~\ref{fig:ue_thr} shows that due to both interference enhancement (see Fig.~\ref{fig:sinr_cdf}) as well as multi-user diversity loss (see Fig.~\ref{fig:cell_ov_thr}),
the average UE throughput for a given number of served UEs drops down with network densifucation.
For instance, keeping 4 UEs per BS and reducing the ISD from 150\,m to 40\,m and 20\,m,
the average UE throughput drops by $\sim$ 42.2\,\% and 59.8\,\%, respectively. 
Comparing PF and  RR performances, 
for an ISD of 20m, the gain of the former is almost negligible,
around 5\,\%.

The minor gains of PF scheduler over the RR one at low ISDs in terms of cell and UE throughput suggests that 
RR scheduler may be a better choice in dense small cell deployments considering the higher complexity of PF scheduler.
This conclusion may have a significant impact in the manufacturing of small cell BSs
where the DSP cycles saved due to the adoption of RR scheduling, can be used to enhance the performance of other technologies. Table~\ref{tab:com} shows the complexity comparison of the discussed scheduling schemes,
where $N^\textrm{UE}$ and $N^\textrm{RB}$ refer to number of UEs and RBs, respectively. 
The complexity of the PF lies in the evaluation of each UE on each RB considering a greedy PF 
that operates on a per RB and subframe basis and is equal to $N^\textrm{UE}\times N^\textrm{RB}$. 
It is worth noting that PF complexity with exhaustive search is considerably higher. 

\begin{figure}[t]
\centering
\includegraphics[scale=0.545]{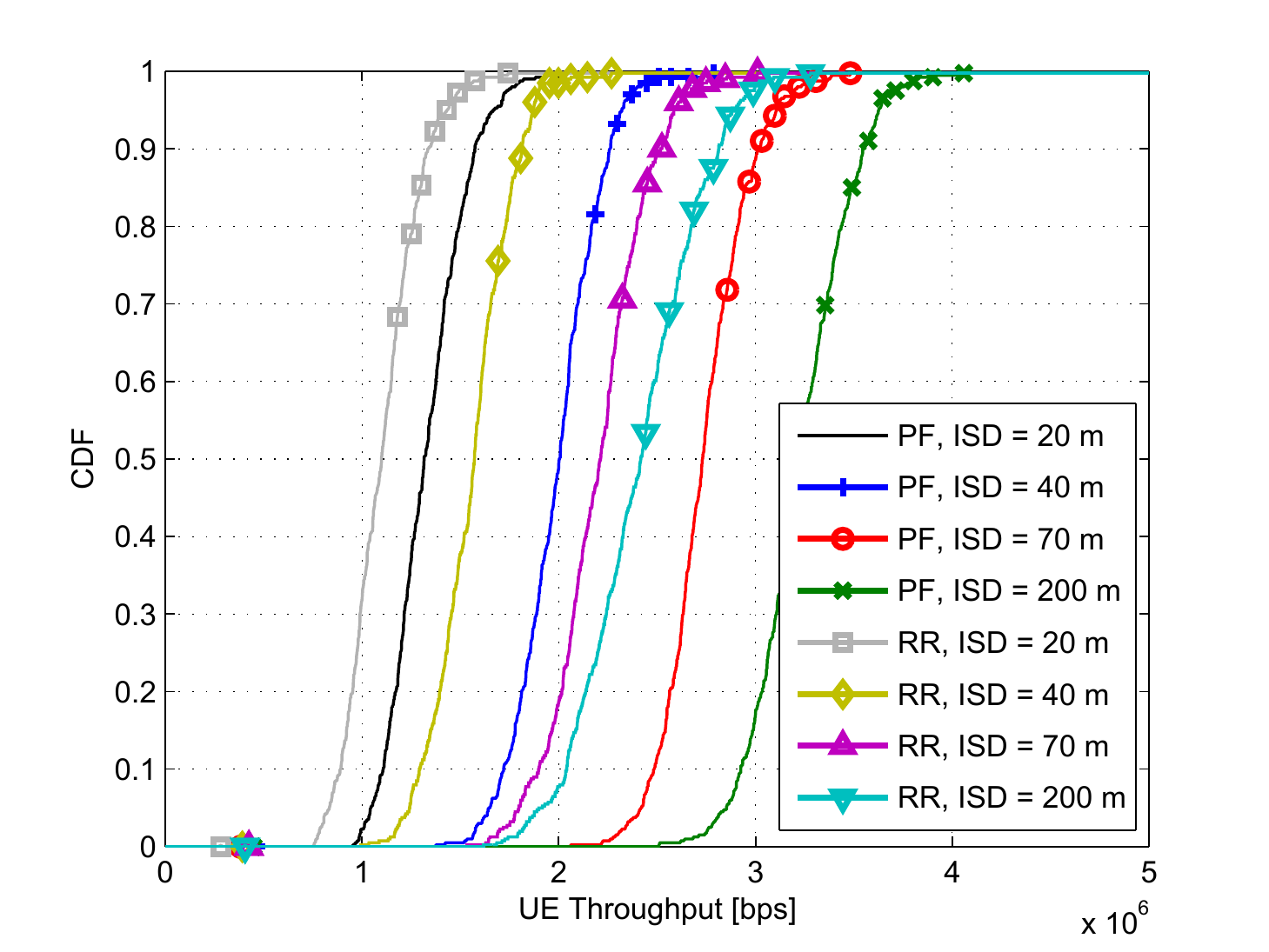}
\caption{CDF of UE throughput for different ISDs.}
\label{fig:ue_cdf}
\vspace{-0.28cm}
\end{figure}

\balance
Until now it has been shown how network densification affects cell and UE performance for a given low number of UEs per BS.
In contrast, it is important to note that shrinking the cell size naturally leads to a lower number of UEs per BS,
which increases UE throughput.
Fig.~\ref{fig:ue_thr} shows that the mean UE throughput is significantly increased by lowering the number of UEs per BS.
At an ISD of 70\,m,
the mean UE throughput is increased by $\sim$ 1.30x, 1.42x and 1.75x when the number of served UEs is lowered from 4 to 3, 2, and 1 UE per cell, respectively.
This UE throughput gain is due to the larger portion of spectrum that each UE can assess,
and is larger than the previously presented UE throughput loss due to the loss in multi-user diversity. Fig.~\ref{fig:ue_cdf} also shows the CDF of the UE throughput for various ISDs for both PF and RR schedulers. It can be realized that reducing the ISD from 200 m to 40 m and 20 m, the UE 5\%-tile throughput drops by 40.8\% and 36.7\%, respectively. Comparing the UE 5\%-tile throughput of PF and RR for an ISD of 20 m, the gain of the former is
almost negligible, around 9\%. This proves how network densification is still worth it from a capacity view point despite both the interference enhancement as well as the multi-user diversity loss.
However, the gains may not be as large as expected due to the mentioned effects. 

\begin{table}
\centering
\renewcommand{\arraystretch}{1.6}
\caption{Comparison of PF and RR schedulers.}
\label{tab:com}
\scalebox{0.85}{
\centering
\begin{tabular}{|>{\centering\arraybackslash}m{1.2cm}|>{\centering\arraybackslash}m{2.4cm}|>{\centering\arraybackslash}m{2.4cm}|>{\centering\arraybackslash}m{1.22cm}|}
\hline \textbf{Scheduling Algorithm} & \textbf{Time Domain} &  \textbf{Frequency Domain} & \textbf{Complexity} \\ \hline
\hline
PF & UE selection according to Eq.~(\ref{eq:td}) & RBs allocation according to Eq.~(\ref{eq:pf}) & $N^\textrm{UE}\times N^\textrm{RB}$     \\ \hline
RR 1 & Single UE (in an iterative mode) & Entire bandwidth & 1 \\ \hline
RR 2 & \textemdash & Bandwidth equally divided among all UEs & 1 \\ \hline
RR 3 & Single UE selection according to Eq.~(\ref{eq:td}) & Entire Bandwidth for single UE & $N^\textrm{UE}$   \\ \hline
RR 4 & Multiple UEs selection according to Eq.~(\ref{eq:td}) & Bandwidth equally divided among specified UEs &  $N^\textrm{UE}$   \\ \hline
\end{tabular}}
\vspace{-0.06cm}
\end{table}

\section{CONCLUSION}
\label{sec:con}
We have discussed that due to LOS propagation,
the Rician fading channel may be more suitable to analyse dense small cell deployments,
and the Rician K factor was derived as a function of distance between the UE and its serving BS.
We have also discussed the dependency of the shadow fading cross-correlation coefficient on the ISD,
and proposed a model for it.
This paper has also shown that as the cell size reduces,
the path loss becomes stronger and so does the interference impact.
As the cell size reduces,
multi-user diversity gains also vanish and serving more UEs does not bring any further gain in cell throughput.
As a major remark, 
the PF scheduling gains over RR are small ($\sim$ 10\%) at low ISDs, 
so RR may be used in dense deployments considering the extra complexity of PF. 
Despite of the increase in interference and the loss in multi-user diversity,
simulations show that network densification is still effective from throughput enhancement view point.

\section*{Acknowledgement}
\label{sec:ack}
The work is partially sponsored by FP7 WiNDOW project.

\bibliographystyle{ieeetr}
\bibliography{references}

\begin{thebibliography}{10}

\bibitem{6815892}
V.~Jungnickel, K.~Manolakis, W.~Zirwas, B.~Panzner, V.~Braun, M.~Lossow,
  M.~Sternad, R.~Apelfrojd, and T.~Svensson, ``The role of small cells,
  coordinated multipoint, and massive mimo in 5g,'' {\em Communications
  Magazine, IEEE}, vol.~52, pp.~44--51, May 2014.

\bibitem{6525591}
I.~Hwang, B.~Song, and S.~Soliman, ``A holistic view on hyper-dense
  heterogeneous and small cell networks,'' {\em Communications Magazine, IEEE},
  vol.~51, pp.~20--27, June 2013.

\bibitem{HetNetbook}
X.~Chu, D.~L\'opez-P\'erez, Y.~Yang, and F.~Gunnarsson, {\em {Heterogeneous
  Cellular Networks: Theory, Simulation and Deployment}}.
\newblock University Cambridge Press, 2013.

\bibitem{3gpp}
``{3GPP TSG RAN, TR 25.996 v10.0.0, ``Spatial Channel Model for Multiple Input
  Multiple Output (MIMO) simulations (release 10)},'' Mar. 2011.

\bibitem{1651489}
H.~Claussen, ``Efficient modelling of channel maps with correlated shadow
  fading in mobile radio systems,'' in {\em Personal, Indoor and Mobile Radio
  Communications, 2005. PIMRC 2005. IEEE 16th International Symposium on},
  vol.~1, pp.~512--516, Sept 2005.

\bibitem{LTE}
F.~Khan, {\em {LTE for 4G Mobile Broadband Air Interface Technologies and
  Performance}}.
\newblock University Cambridge Press, 2009.

\bibitem{4063519}
M.~Sharif and B.~Hassibi, ``A comparison of time-sharing, dpc, and beamforming
  for mimo broadcast channels with many users,'' {\em Communications, IEEE
  Transactions on}, vol.~55, pp.~11--15, Jan 2007.

\bibitem{4146798}
T.~Bu, L.~Li, and R.~Ramjee, ``Generalized proportional fair scheduling in
  third generation wireless data networks,'' in {\em INFOCOM 2006. 25th IEEE
  International Conference on Computer Communications. Proceedings}, pp.~1--12,
  April 2006.

\bibitem{4489366}
E.~Liu and K.~Leung, ``Proportional fair scheduling: Analytical insight under
  rayleigh fading environment,'' in {\em Wireless Communications and Networking
  Conference, 2008. WCNC 2008. IEEE}, pp.~1883--1888, March 2008.

\bibitem{4526113}
G.~Monghal, K.~Pedersen, I.~Kovacs, and P.~Mogensen, ``Qos oriented time and
  frequency domain packet schedulers for the utran long term evolution,'' in
  {\em Vehicular Technology Conference, 2008. VTC Spring 2008. IEEE},
  pp.~2532--2536, May 2008.

\bibitem{851593}
A.~Jalali, R.~Padovani, and R.~Pankaj, ``Data throughput of cdma-hdr a high
  efficiency-high data rate personal communication wireless system,'' in {\em
  Vehicular Technology Conference Proceedings, 2000. VTC 2000-Spring Tokyo.
  2000 IEEE 51st}, vol.~3, pp.~1854--1858 vol.3, 2000.

\bibitem{6362524}
H.~Claussen and L.~Ho, ``Multi-carrier cell structures with angular offset,''
  in {\em Personal Indoor and Mobile Radio Communications (PIMRC), 2012 IEEE
  23rd International Symposium on}, pp.~1179--1184, Sept 2012.

\bibitem{4525984}
P.~Skillermark, M.~Almgren, D.~Astely, M.~Lundevall, and M.~Olsson,
  ``Simplified interference modeling in multi-cell multi-antenna radio network
  simulations,'' in {\em Vehicular Technology Conference, 2008. VTC Spring
  2008. IEEE}, pp.~1886--1890, May 2008.

\end{thebibliography}

\end{document}